\documentclass[%
 reprint,
 amsmath,amssymb,
 aps, nofootinbib
]{revtex4-1}
\usepackage{color}
\usepackage{graphicx}
\usepackage{dcolumn}
\usepackage{bm}
\usepackage{hyperref}

\begin{document}

\preprint{APS/123-QED}

\title{A complete model of the CR spectrum and composition \\across the Galactic to Extragalactic transition}

\author{Noemie Globus}
\affiliation{School of Physics \& Astronomy, Tel Aviv University, Tel Aviv 69978, Israel}%
\author{Denis Allard}
\author{Etienne Parizot}
\affiliation{Laboratoire Astroparticule et Cosmologie, Universit\'e Paris Diderot/CNRS, 10 rue A. Domon et L. Duquet, 75205 Paris Cedex 13, France}

%
%
%
%

\begin{abstract}
We present a complete phenomenological model accounting for the evolution of the cosmic-ray spectrum and composition with energy, based on the available data over the entire spectrum. We show that there is no need to postulate any additional component, other than one single Galactic component depending on rigidity alone, and one extragalactic component, whose characteristics are similar to those derived from a study of particle acceleration at mildly relativistic shocks in a GRB environment (Globus et al., 2015). In particular, we show that the resulting cosmic ray spectrum and composition satisfy the various constraints derived from the current data in the Galactic/extragalactic transition region, notably from the measurements of KASCADE Grande and Auger. Finally, we derive some generic features that a working phenomenological scenario may exhibit to give a global account of the cosmic ray data with a minimum number of free parameters. 
\end{abstract}

\maketitle


\section{\label{sec:Introduction}Introduction}

One century after the discovery of cosmic rays (CRs), their study remains one of the main focus of high energy astrophysics and astroparticle physics. In the recent years, an important set of new cosmic-ray data have become accessible, thanks to major experimental progresses. In particular, the Pierre Auger Observatory (Auger)\cite{Auger04} and Telescope Array\cite{TA} experiments have explored the ultra-high-energy cosmic rays (UHECRs) with unprecedented observational power. Below the ankle, KASCADE-Grande (KG)\cite{KG10} has made crucial measurements of the cosmic-ray spectrum, including a separation between low-mass and high-mass nuclei, which can be linked to measurements in the knee region made in particular by the KASCADE\cite{KASCADE} experiment. Below the knee, new data have been available as well, notably from CREAM\cite{CREAM04} and TRACER\cite{TRACER07} and most recently from PAMELA\cite{PAMELA13} and AMS\cite{AMS13}, allowing a more precise determination of the relative abundances of the various nuclei among the Galactic cosmic rays (GCRs). 

These new data allow a more complete description of the CR phenomenology over the entire energy range and in particular at the Galactic/extragalactic (GCR/EGCR) transition, taking into consideration both the CR energy spectrum and the composition. Some recent studies have claimed that the data could not be accounted for without invoking, in addition to the main GCR and extragalactic UHECRs, one or more additional (Galactic or extragalactic) components dominating the flux at intermediate energies, between the knee and the ankle (see for instance \cite{Hillas05, Tunka2013, Aloisio14, Giacinti15}), and whose spectrum, composition and flux normalisation could be adjusted at will to reconcile the observations with one's preconception regarding the low-energy and/or high-energy CRs. In this Communication, we argue against such a necessity, and explicitly show that the global CR spectrum can be fully described in a natural way within a two-component model, one Galactic and one extragalactic, without the need of introducing a new, hypothetical component.

We base our model on a very simple and generic description of the GCR component, whose properties depend only on the rigidity of the particles (that is, for ultra-relativistic particles, the energy-to-charge ratio), and on an EGCR component that is directly borrowed from a previous work on particle acceleration at mildly relativistic internal shocks of GRBs\cite{Globus14}, which consistently predicts an evolution of the composition compatible with Auger data. We show that this model provides a fair account of all the significant features of the cosmic ray spectrum and composition from below the knee to the highest energies.

\section{\label{sec:model}Model}

\subsection{The extragalactic component}

One of the most important results of Auger is the observation of a transition from a light-dominated to a heavy-dominated composition between a few $10^{18}$~eV and a few $10^{19}$~eV \cite{Aab14a, Aab14b}. As we first proposed in \cite{Allard08}, this can be interpreted 
as a consequence of a low-energy cutoff of the protons in the sources, due to an intrinsic limitation of the acceleration process. As long as the acceleration of particles is governed only by electromagnetic processes, all nuclei behave in exactly the same way if they have the same magnetic rigidity. Nuclei of charge $Z$ are thus expected to reach an energy $E_{\max}(Z) = Z\times E_{\max}(p)$, where $E_{\max}(p)$ is the maximum proton energy. It is thus natural to expect that nuclei of higher and higher mass dominate the UHECR source composition above the maximum energy of the lower-mass nuclei.
\begin{figure*}[t!]
\includegraphics[width=1.\linewidth]{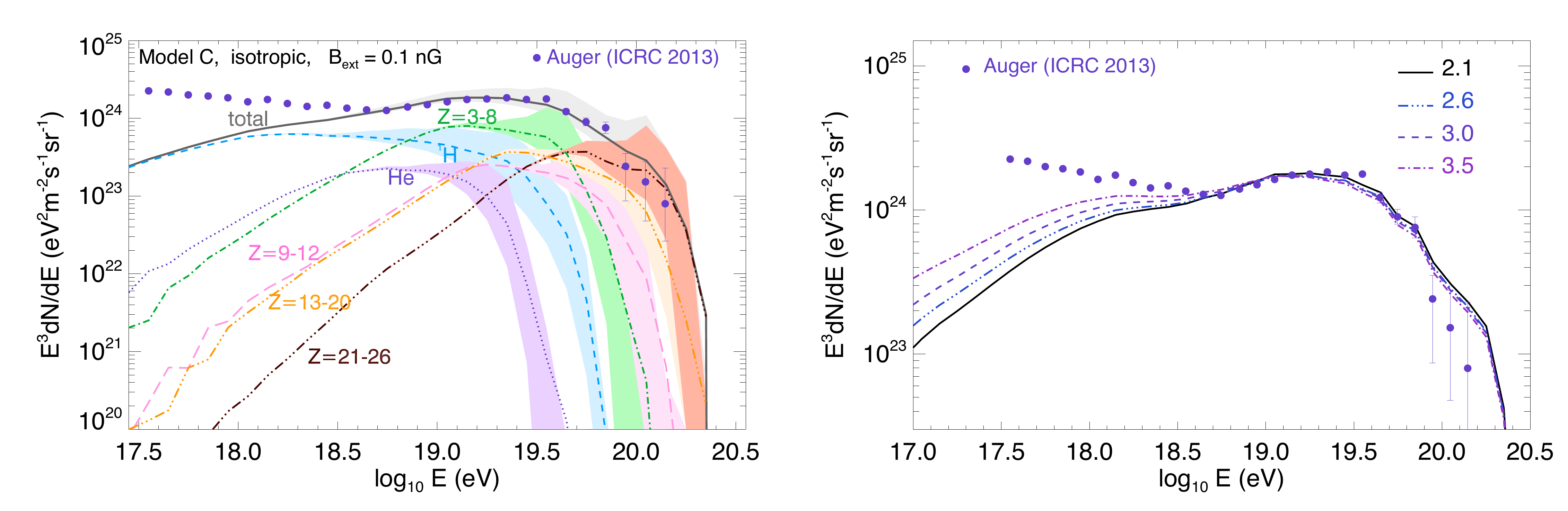} 
\caption{\label{fig:diffusC} Left: EGCR flux as a function of energy for H, He and different ranges of nuclei (indicated by the charge labels), as predicted by our acceleration model in GRBs, adjusted to the Auger data. The figure was taken from \cite{Globus14}. Right: Effect on the propagated EGCR spectrum of different assumptions for the cosmological evolution of the source density, in $(1 + z)^{\alpha}$, with $\alpha = 2.1$, 2.6, 3.0 or 3.5, as indicated.}
\end{figure*}
In more realistic situations, however, such a simple model should be amended to take into account the energy losses and photo-dissociation processes that may occur in the acceleration site. In a recent study, we developed a numerical model for the acceleration of UHECRs in the mildly relativistic internal shocks of a GRB \cite{Globus14}. 
We showed that the relatively high density of energetic photons in the acceleration site leads to significant photo-dissociation, which has two important consequences in the present context: i) the resulting maximum energy of the nuclei is not strictly proportional to $Z$, but also reflects their photo-dissociation rate; and ii) the spectrum of the UHECRs eventually injected by the source into the intergalactic medium is close to a hard power-law (roughly in $E^{-1}$ below $E_{\max}$), but while all composed nuclei have essentially the same spectral index, protons have a significantly steeper spectrum. This is due to the secondary neutrons, which are  mostly produced by photodisintegration processes during the acceleration. Indeed, the charged particles mostly escape from the acceleration region in the weak scattering regime, i.e. at the highest energies. 
On the contrary, the secondary neutrons are not confined by the local magnetic fields, and thus escape with their production spectrum (which is similar to that of the nuclei at the shock), flowing freely out of the source before decaying into protons (see \cite{Globus14} for more details).

The model consistently predicts the shape of the spectra of individual nuclei, including their high-energy cutoff at the source. We then convoluted individual source injection over the GRB luminosity function and used our UHECR propagation code, taking into account energy losses, photo-dissociation and magnetic deflections\cite{Globus08}, to derive the propagated spectrum which can be observed on Earth. The result is shown in Fig.~\ref{fig:diffusC}, which is taken from \cite{Globus14}, with a comparison to the Auger data. The shaded area corresponds to the so-called cosmic variance and represent the expected range for the flux of the different nuclei, including 90\% of independent realisations of the model. As can be seen, the propagated proton spectrum is indeed much softer than that of the other nuclei. This model reproduces fairly well the overall spectrum, and shows a clear transition from a proton-dominated composition at the ankle to a Fe-dominated composition at the highest energies.

In order to limit the number of free parameters, we use this model of the EGCR component in the present study. However, we allow for different assumptions regarding the cosmological evolution of the sources. Figure~\ref{fig:diffusC} shows the total spectrum obtained when the source density increases as function of redshift as $(1 + z)^{\alpha}$, with $\alpha = 2.1$, 2.6, 3.0 or 3.5. As expected, changing the cosmological evolution does not affect the high-energy part of the spectrum, since the contributing sources are all located at low redshifts, due to the GZK horizon effect. However, a stronger source evolution implies a larger contribution of the EGCR sources at low energy. Since the corresponding flux is dominated by protons, larger values of $\alpha$ result in larger contributions of extragalactic protons, which influences the composition at the GCR/EGCR transition. From our calculations, we found that a relatively large evolution, with $\alpha$ between 3.0 and 3.5, provides the most striking agreement with the composition measurements over the whole energy range. In the following, we use $\alpha = 3.5$, which is fully compatible with the observational constraints. In particular we verified that the gamma-ray emission resulting from the intergalactic showers associated with the propagation of the EGCRs does not violate the measurements made by the Fermi telescope.  We should however note that our results do not necessarily imply that the EGCR sources must have a strong cosmological evolution. 
Similar changes of the EGCR spectrum could also be obtained for instance by modifying the assumed source luminosity function or the turbulence structure at the shock. Moreover, the assumption that GRBs are the sources of UHECRs is not critical for the success of the model. The key feature on which we rely here is the prediction of a softer spectrum for the protons, which can be expected in other cosmic accelerators as soon as they involve a significant amount of nuclei and the matter or radiation density is large enough in the source environment.

\subsection{The Galactic component}

For the GCR component, we assume that all nuclei have the same rigidity spectrum, consisting of a broken power law with spectral index $x$ below an energy $E_{\mathrm{break}}(Z) = Z\times E_{\mathrm{break}}(\mathrm{p})$, and a spectral index $x + \Delta x$ above that energy, up to an exponential cutoff in $\exp(E/ZE_{\max})$ above $E_{\max}$. 

The slope $x$ and the relative abundance of the various nuclei  are simply adjusted on the most recent available data at low energy\cite{Maurin2013}. Note that, we only use data above 300 GV,  given the evidence found in PAMELA data\cite{PAMELA11} of a change of slope of the spectrum of H and He below this rigidity (also confirmed by the most recent AMS data\cite{AMS15}). Above 300 GV, we found that an index $x = 2.67$ provides a good fit of the data for all nuclei. In particular, TRACER\cite{TracerPropa, Tracer03, Tracer06} finds an index of $x = 2.67 \pm 0.08$ for the combined elements heavier than He, while CREAM\cite{CREAM1} find an index of $2.66 \pm 0.02$ protons, and $2.66 \pm 0.04$ for elements heavier than He\cite{CREAM2}. There is admittedly some tension with the measurements of CREAM in the specific case of He nuclei\cite{CREAM1}, namely $x(\mathrm{He}) = 2.58\pm0.02$. However, the most recent AMS02\cite{AMS15} results are in good agreement with $x(\mathrm{He})=2.67$. Therefore, instead of leaving the spectral index free for each nucleus, we decided to stick to the most natural assumption that all nuclei have the same spectrum in rigidity, and simply determine the relative abundances from the observational data. A combined fit of CREAM and AMS02 data is used to normalise the proton flux while only AMS02 is used for He nuclei, and a combined fit of CREAM, TRACER and ATIC-2\cite{Atic02} data (whenever available) is used for heavier nuclei (CNO, Ne, Mg, Si, S, Ar, Ca and Fe). While some small corrections might be due to spallative processes (see e.g. \cite{TracerPropa}), we note that the model is anyway robust with respect to the assumption of a single index $x$ for all elements, since a change of 0.03 in the value of $x$ for an element would result in a mere 23\% difference of its relative abundance after three decades in energy, which we verified has very minor impact on our results.

The change of slope $\Delta x$ (the knee) is a major feature in the GCR spectrum; its origin may be on the side of the acceleration process, for instance a consequence of a reduction of the number of sources contributing at higher and higher energy (see \cite{Drury12} and references therein). Or it may be a feature of the propagation of the GCRs in the interstellar medium, through a change of the diffusion regime, or the effect of a Galactic wind (e.g. \cite{Ptuskin12} and references therein; see also \cite{Giacinti15} for a recent account). It could also reflect some inhomogeneity in the GCR flux, associated with the granularity and/or intermittency of the sources.  In the current modelling, we do not attempt to give any interpretation of this change of slope. $\Delta x$ is adjusted in order to obtain a good fit of the spectrum of individual nuclei (or classes of nuclei), as given by KASCADE\cite{Antoni05, Finger11} and KASCADE-Grande\cite{Apel11, Apel13, Bertaina13, Apel14}. We find values of $\Delta x$ between $\sim$0.3 and 0.5 coupled to values of  $E_{\mathrm{break}}(p)$ between $\sim2\times 10^{15}$ and $4\times10^{15}$~eV .

Finally, the only remaining free parameter of the GCR component model is the energy scale, $E_{\max}$, of the high-energy cutoff, which we adjust to ensure a good fit of the data in the GCR/EGCR region. In particular values of $E_{\max}$ between $\sim5\times10^{16}$ and $1.5\times10^{17}$~eV allow the heavy Galactic component to become very low to negligible at the ankle (which thus marks roughly the end of the GCR/EGCR transition \cite{Allard05, Allard07}) and the combined light Galactic and extragalactic component to produce an ankle around $\sim 10^{17}$~eV, as observed in KG data \cite{Apel13, Bertaina13}.
\begin{figure*}[t!]
\includegraphics[width=1.\linewidth]{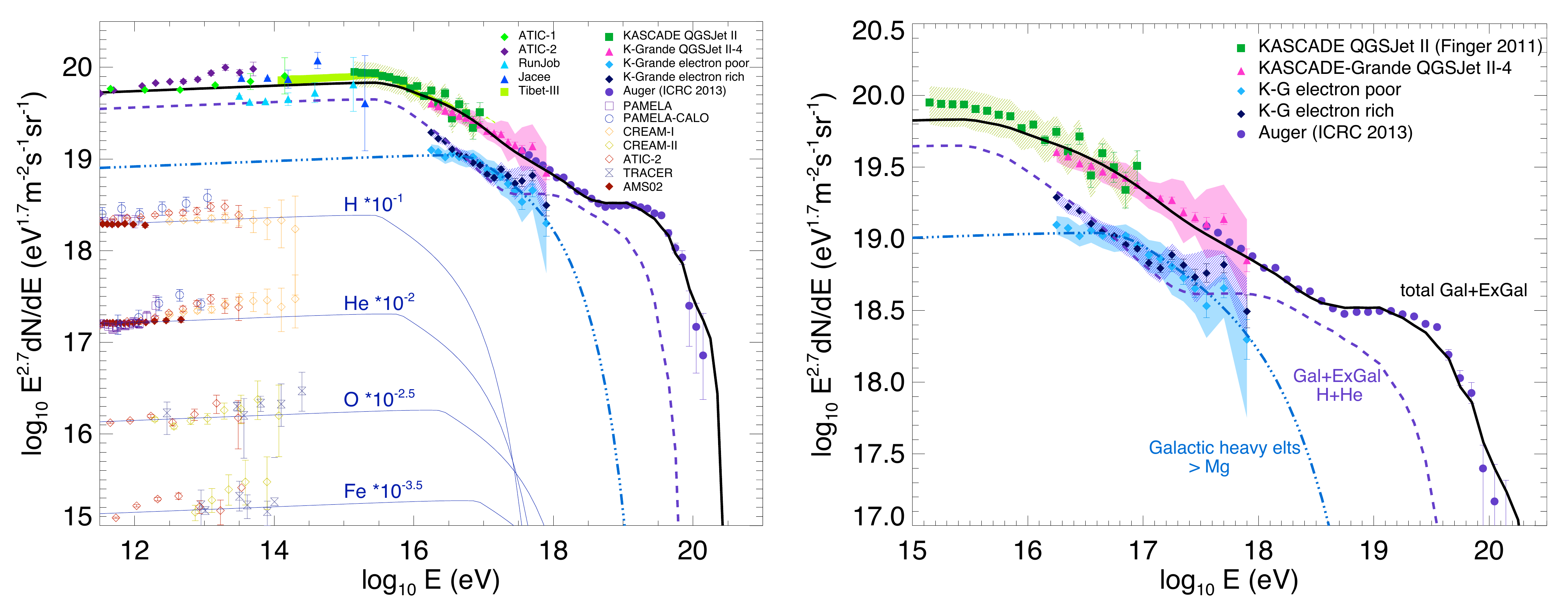}
\caption{\label{fig:GlobalSpectrum} Comparison between the spectrum of our global CR model and the available data, detailed in the upper right corners. The individual spectra of H, He, O and Fe are shown for the GCR component (left, with the indicated rescaling). In addition, the ``heavy'' and ``light'' components determined by KASCADE-Grande are shown with dashed and dotted-dashed line (see text). Right: close up view, showing the sum of the H and He fluxes (dashed) and the elements heavier than Mg (dotted-dashed).}
\end{figure*}

\section{\label{sec:results}Results}

In this section, we show the results obtained with the above model, after setting the values of the parameters to $E_{\mathrm{break}}(p)=10^{15.5}$~eV, $\Delta x=0.45$ and $E_{\max}=6\times10^{16}$~eV, and combining the resulting Galactic component to the EGCR model using $\alpha=3.5$.

The resulting spectrum is shown in Fig.~\ref{fig:GlobalSpectrum}. The left panel shows a global view of the CR {\it energy} spectrum above $10^{11.5}$~eV. In particular the assumed spectra of the 4 most abundant Galactic  species (H, He, O and Fe) are compared with satellite and balloon borne measurements from various experiments (CREAM\cite{CREAM1, CREAM2}, PAMELA{\cite{PAMELA11}, PAMELA-CALO\cite{PAMELA-CALO}, ATIC-2\cite{Atic02} and TRACER\cite{Tracer03, Tracer06}), while the total spectrum is compared below the knee with various balloon borne or ground based experiments (RUNJOB\cite{Runjob}, JACEE\cite{Jacee}, ATIC-1\cite{Atic01}, ATIC-2\cite{Atic02} and Tibet-III\cite{TibetIII}}. Taking into account the existence of some discrepancies between the various experimental results, our Galactic component is in good agreement with data for both the total spectrum and the different components below the knee energy. The right panel shows a close up view at higher energy including the knee, the ankle and the highest energy regions. Our model predictions are compared with KASCADE\cite{Finger11} as reconstructed with the QGSJetII-3 model\cite{QGSJet-II}, KG\cite{Bertaina13} as reconstructed with QGSJetII-4\cite{QGSJet-II4} and Auger\cite{Aab13} data.
The dashed line shows the flux obtained by summing the H and He spectra, including both GCRs and EGCRs. This is in good agreement with the KG data, when only the so-called ``light'' component (corresponding to electron-poor showers) is taken into account, which is intended to correspond to H and He nuclei according to the analysis presented in \cite{Apel13, Bertaina13}. The dotted-dashed line represents the flux of all the elements heavier than Mg, and is to be compared with the so-called ``heavy'' component. As can be seen, both components are consistent with the data, and fit well within the KG systematic errors. In particular, we find that the heavy component exhibits a knee slightly below $10^{17}$~eV, while the light component shows an ankle slightly above that energy as the light EGCR component become dominant over the sharply decreasing light Galactic component. The behaviors of both the heavy and light components are in very good agreement with the KG findings\cite{Apel11, Apel13, Bertaina13, Apel14}.

Of course, the quantitative agreement between our model and the data depends on the underlying assumption regarding the hadronic model used to reconstruct KG data. However, the choice of the hadronic model mostly influences the relative normalization of the different components. In particular, for all the hadronic models tested in \cite{Bertaina13, Apel14}, the heavy knee and light ankle features remain, and the post knee shapes of the light and heavy component remain similar to one another. Moreover, we found that the relative abundances for these components predicted by our model agree best with the data reconstructed with the QGSJetII-4 which is, to date, the most recent of the models tested against KG data and the only one that includes recent experimental constraints for LHC data\footnote{EPOS-LHC, the most recent version of the EPOS\cite{EPOS1, EPOS2} model (including constraints from LHC data) has not yet been confronted to KG data. Given the quoted similarity between EPOS-LHC and QGSJetII-4 in terms of the predicted correlation between the shower size and the muon number in KG energy range, one can however anticipate that the light and heavy components deduced from KG data should be quite similar for both models. Likewise these models should yield a lighter composition at the knee than previously estimated from KASCADE data\cite{Antoni05, Finger11}, probably more consistent with direct measurements at lower energies.}.

\begin{figure*}[t!]
\includegraphics[width=1.\linewidth]{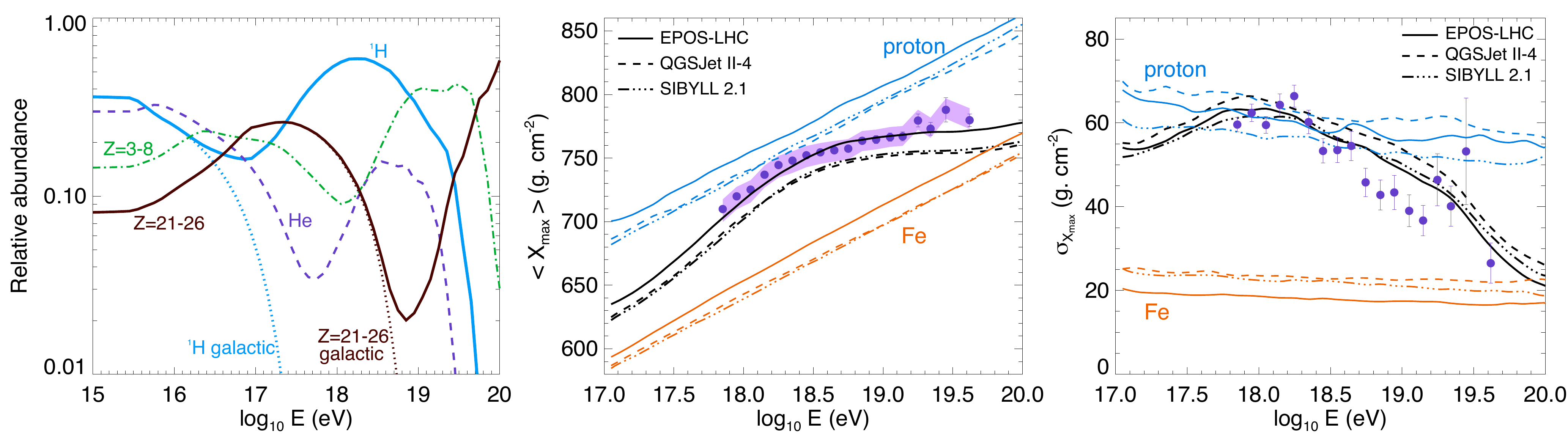} 
\caption{\label{fig:composition} Left: Relative abundance of H, He and the elements in the indicated charge ranges, as a function of energy. For H and the elements between Sc and Fe, the purely Galactic component is shown with faint dotted lines. Center: comparison between the model predictions for the evolution with energy of the depth of the shower maximum, $X_{\max}$, and the Auger data, for three different hadronic models. Right: same as central panel, for the $X_{\max}$ variance, $\sigma(X_{\max})$.}
\end{figure*}

In addition to these spectral features, our model provides a detailed description of the CR composition from the knee to the highest energies, which allows a comparison with the data. The left panel of Fig.~\ref{fig:composition} shows the relative abundance of H, He and the following two dominant sets of nuclei, namely CNO (and sub-CNO) nuclei, and sub-Fe (and Fe) nuclei. For protons and sub-Fe nuclei, we also show separately the Galactic component, using dotted lines.

As can be seen, even though the Galactic protons essentially disappear at $\sim1.5\,10^{17}$~eV, the abundance of protons never drops below 15\%, and rises up again to more than 50\% (with a maximum around 60\%) just above $5\,10^{17}$~eV. The fact that these protons, which ensure a dominantly light component across the ankle, are extragalactic protons, is fully consistent with the anisotropy measurement of Auger. Indeed, a Galactic component of protons would most probably produce a significant anisotropy towards the Galactic center and/or plane, which is excluded by the data. Finally, the proton fraction is seen to rapidly decrease above the ankle, to finally vanish above a few $10^{19}$~eV, letting heavier and heavier nuclei dominate the UHECR spectrum.

The behaviour of Fe (and sub-Fe) nuclei is quite different, as there is practically no overlap between the Galactic component, which ends at a few $10^{18}$~eV (i.e. 26 times higher in energy than the Galactic protons), and the extragalactic component, which rises up strongly above $10^{19}$~eV, to reach 60\% at $10^{20}$~eV.

It is interesting to note that, according to our model, the dominant class of nuclei over roughly one decade in energy, between $\sim 6\,10^{18}$~eV and $\sim 5\,10^{19}$~eV, should in fact be CNO. This appears in very good agreement with the recent Auger findings \cite{Aab14b}.

The spectra of individual nuclei are unfortunately very difficult to measure separately, which prevents a direct comparison with the data. However, it is possible to compare the data with the model predictions for the composition-dependent observables, namely the depth of the maximum shower development, traditionally referred to as $X_{\max}$, and its spread (among the whole set of showers) at a given energy, $\sigma(X_{\max})$.
This is done in 
Fig.~\ref{fig:composition}, where we plotted the evolution of these two observables (central and right panels) with energy, together with the Auger data. For this, we simulated the development of a large number of cosmic-ray showers for the different nuclei and energies, using the CONEX shower simulator\cite{Conex} with three different choices of the hadronic interaction model (SIBYLL2.1\cite{SIBYLL}, QGSJetII-4\cite{QGSJet-II4} and EPOS-LHC\cite{EPOS1, EPOS2}. The agreement between the prediction of our model and the data is remarkable over the entire energy range, both qualitatively and quantitatively, especially when the shower development is calculated using the EPOS-LHC hadronic model. It is again interesting to note that this model takes into account the recent constraints from measurements performed at LHC. Although they probably do not reproduce perfectly all air showers properties\cite{Aab15}, the most recent hadronic models seem to give a more coherent picture of the evolution of the composition deduced from indirect measurements, from the knee to the highest energies.

\section{\label{sec:Summary}Summary}

We showed that the whole CR spectrum, including the key region of the GCR/EGCR transition, can be described by simply superposing a rigidity dependent GCR component and a generic EGCR model, without additional degrees of freedom.

In our model, the GCR component is identical for all nuclei with the same rigidity. The maximum energy of protons accelerated in Galactic sources is $\sim 6\,10^{16}$~eV, and the transition towards extragalactic protons takes place around $10^{17}$~eV, where KASCADE-Grande observes an ankle in the light CR component. While the knee-like break in the GCR proton component occurs at $\sim 3\,10^{15}$~eV, the corresponding break in the Fe components appears at $\sim 8\,10^{16}$~eV, which is in agreement with the observed ``heavy-knee'' in the KASCADE-Grande data. The normalisations of the light and heavy components are also in good agreement with the data.

Our results suggest that extragalactic protons account for more than 50\% of the total flux from $\sim5\,10^{17}$~eV to $\sim5\,10^{18}$~eV, and drop below 10\% above $3\,10^{19}$~eV. The dominant class of nuclei between $\sim 6\,10^{18}$~eV and $\sim 5\,10^{19}$~eV is CNO. The evolution of the composition predicted by our model has been shown to be fully compatible with the Auger data\cite{Aab14a,Aab14b}, across the observed transition from a light-dominated to a heavy-dominated composition. 


An important reason for the success of the model is the fact that the EGCR source spectrum is significantly steeper for protons than for the heavier nuclei. As recalled above, this is because most of EGCR protons injected in the intergalactic medium below $\sim 10^{19}$~eV, are in fact decay products of freely escaping secondary neutrons, produced during the acceleration through the photo-dissociation of heavier nuclei. While this is a direct consequence of our particle acceleration model, presented in detail in \cite{Globus14}, we believe that it is a generic feature of UHECR acceleration processes occurring in photon-rich environments. As a consequence, the agreement between the model predictions and the data should not be taken \emph{per se} as an argument in favour of the GRB source model, but in favour of the global CR interpretation scheme developed in this study. Combined with a relatively low maximum proton energy, a single EGCR component can thus explain in a consistent way the evolution of the composition above $10^{18}$~eV as well as the spectral and composition features observed below the ankle, down to the knee. In particular, the relative energy scale between the ``light ankle''  and the ankle (a factor $\sim 30 \sim Z(\mathrm{Fe})/Z(\mathrm{p})$) is most naturally explained within this framework involving a transition between two components.

We thus conclude that, unless new data will contradict the current observational status, there is no need for invoking any additional CR component. Now, if one opts indeed for the minimal assumption that a single component accounts for all GCRs, then our results strongly suggest that this component should be able to accelerate protons up to at least $\sim5-6\, 10^{16}$~eV. This might be in tension with the generally accepted models for particle acceleration at the shocks of individual supernova remnants in the Galaxy, which hardly reach energies much larger than $10^{15}$~eV. This could suggest that these models should be modified, or that other classes of sources may be have a dominant contribution to the GCRs, perhaps through collective acceleration processes as might be expected in superbubbles (see e.g. \cite{Parizot14} and references therein).

\begin{acknowledgments}
We wish to thank Mario Bertaina and Eun-Joo Sein Ahn for very helpful discussions. 
NG acknowledges a grant from the Israel Science Foundation no. 1277/13.
\end{acknowledgments}


\end{document}